\documentclass[10pt,a4paper]{article}
\usepackage{jheppub_kim}
\usepackage{pdflscape}
\usepackage{amsmath}
\usepackage{amssymb}
\usepackage{dcolumn}
\usepackage{bm}
\usepackage{color}
\usepackage{epsfig}
\usepackage{amsfonts}
\usepackage{graphicx}
\usepackage{subfigure}
\usepackage{dcolumn}
\begin{document}
\title{Quantum  Thermodynamics of an M2-M5 Brane System}
\author[a]{Behnam Pourhassan,}
\author[b]{Houcine Aounallah,}
\author[c]{Mir Faizal,}
\author[d]{Sudhaker Upadhyay,}
\author[e]{Saheb Soroushfar,}
\author[f]{Yermek O. Aitenov,}
\author[g]{Salman Sajad Wani.}
\affiliation[a] {School of Physics, Damghan University, Damghan, 3671641167, Iran.}
\affiliation[b] {Department of Science and Technology. Larbi Tebessi University, 12000 Tebessa, Algeria.}
\affiliation[c] {Department of Physics and Astronomy, University of Lethbridge, Lethbridge, Alberta, T1K 3M4, Canada.}
\affiliation[c] {Irving K. Barber School of Arts and Sciences, University of British Columbia, Kelowna, British Columbia, V1V 1V7, Canada.}
\affiliation[c] {Canadian Quantum Research Center 204-3002 32 Ave Vernon, BC V1T 2L7 Canada.}
\affiliation[d] {Department of Physics, K.L.S. College,  Nawada-805110, constituent unit of Magadh University, Bodh-Gaya), Bihar, India.}
\affiliation[d] {Visiting Associate, Inter-University Centre for Astronomy and Astrophysics (IUCAA), Pune-411007, Maharashtra, India.}
\affiliation[e] {Faculty of Technology and Mining, Yasouj University, Choram 75761-59836, Iran.}
\affiliation[f] { Department of Genetics, University of Cambridge, Downing Street, Cambridge CB2 3EH, United Kingdom.}
\affiliation[g] {Canadian Quantum Research Center 204-3002 32 Ave Vernon, BC V1T 2L7 Canada.}

\emailAdd{b.pourhassan@du.ac.ir}
\emailAdd{houcine.aounallah@univ-tebessa.dz}
\emailAdd{mirfaizalmir@gmail.com}
\emailAdd{sudhakerupadhyay@gmail.com}
\emailAdd{soroush@yu.ac.ir}

\abstract{
We will investigate a system of M2-M5 branes as a black M2-M5 bound state. The behavior of this system will be investigated at short distances. At such scales, we will have to incorporate quantum gravitational corrections to the supergravity solutions. We will study the non-equilibrium quantum thermodynamics of this black M2-M5 bound state. The quantum work for this solution will be obtained using the Jarzynski equality. We will also study the corrections to the thermodynamic stability of this system from quantum gravitational corrections. We will use the concept of a novel quantum mass to analyze the quantum gravitational corrections to the information geometry of this system. This will be done using effective quantum metrics for this system. }

\maketitle

\section{Introduction}
The standard formalism for black hole thermodynamics has been used to study the thermodynamics of black brane solutions \cite{bb12, bb14}. It is also possible to study the thermodynamics of a system of black branes using this blackfold approach \cite{bf1, bf2, bf4, bf6}.
Here the dynamics of such a branes system can be replaced by an effective worldvolume theory \cite{bf, bf7}.
Such an approach has been used to construct a BIon \cite{b12, b14}.
This BIon describe a wormhole
with an F-string charge connecting a D3-brane to a parallel
anti-D3-brane. It can also be described by F1-D3 intersection \cite{b12, b14}. It is possible to head up this BIon, and investigate the thermodynamics of this finite temperature BIon \cite{BIon01, bion}. The F1-D3 intersection is U-dual to a system of M2-M5 branes \cite{BIon01, bion}. This has motivated the study of the thermodynamics of such a black M2-M5 brane system \cite{bf1}.
The dynamics of two M2-branes are described by the
BLG theory \cite{blg, blg1} and the dynamics of multiple M2-branes are described by ABJM theory \cite{abjm}. These theories are superconformal field theories. The BLG theory is a $\mathcal{N} = 8$ superconformal field theory \cite{blg, blg1}. Even though the ABJM theory only has $\mathcal{N} = 6$ supersymmetry, it is can be enhanced to full $\mathcal{N} = 8$ supersymmetry \cite{abjm1}. It is also expected that M5-branes will be described by a $(2,0)$ superconformal field theory \cite{m51}. It has also been possible to study M2-M5 intersection \cite{m5super}. The superconformal field theories associated with M2 branes ending on M5 branes have been studied using both the ABJM theory \cite{wzw1} and BLG theory \cite{wzw2}.
It is expected that the equilibrium thermodynamics of black branes will be corrected by quantum gravitational corrections at short distances. It has been demonstrated that the holographic principle holds at short distances even after the effects of quantum gravitational corrections have been considered \cite{4a, 5a, 6a, 7a}. In fact, such quantum gravitational corrections to the entropy of black holes have been obtained using AdS/CFT correspondence, which is a concrete realization of the holographic principle \cite{18, 18a, 18b, 18c, 18d}.

It is possible to use the density of microstates associated with conformal blocks of a conformal field theory to obtain corrections to the entropy of a black hole \cite{Ashtekar}. In fact, the corrections to the Cardy formula of a conformal field theory have been used to obtain corrections to the entropy of a black hole \cite{Govindarajan}. The Rademacher expansion can also be used to obtain corrections to the entropy of a black hole \cite{29}. The quantum gravitational corrections to the thermodynamics of black holes have also been investigated using the extremal limit of black holes \cite{19, 19a}.
So, several different approaches have been used to investigate quantum gravitational corrections to the entropy of black holes.
It has been argued that such quantum gravitational corrections to the thermodynamics of black holes can be expressed as a function of the original entropy \cite{2007.15401}. It is possible to use Kloosterman sums to obtain these quantum gravitational corrections to the thermodynamics of black holes \cite{Dabholkar}. Such quantum gravitational corrections can also be obtained using the field theoretical dual to the near horizon geometry \cite{ds12, ds14}.

It is also possible to obtain the corrected entropy of a black hole by analyzing the thermal fluctuations to their thermodynamics \cite{32, 32a, 32b, 32c, 32d}.
In the Jacobson formalism, the geometry of spacetime is viewed as an emergent structure, which emerges from thermodynamics \cite{gr12}. Thus, it is possible to use the Jacobson formalism to investigate the quantum fluctuation to the geometry of spacetime using the thermal fluctuations \cite{gr14}. Thus, motivated by Jacobson formalism, the thermal fluctuations have been used to analyze quantum fluctuations of various black hole solutions \cite{40a, 40b, 40c, 40d}. These thermal fluctuations can be neglected for large black holes because the temperature of black holes scales inversely with their size. Furthermore, for such large black holes we can also neglect the quantum fluctuations as they can be obtained from these thermal fluctuations \cite{gr12, gr14}. Now, these fluctuations cannot be neglected on sufficiently small scales, and it is important to study their effects on black holes \cite{40a, 40b, 40c, 40d}. Here the corrections are expressed using the original equilibrium entropy and temperature of black holes \cite{32, 32a, 32b, 32c, 32d}.

However, the equilibrium description cannot be used to analyze the system, when the size of the black holes is comparable to the Planck scale. However, it is known that to study thermodynamics systems at quantum scales, we have to use non-equilibrium quantum thermodynamics \cite{adba0, adba1, adba2, adba4}.
This has motivated the study of non-equilibrium quantum thermodynamics of black holes \cite{rz12, rz14}.
It may be noted that quantum gravitational corrections to the
original Bekenstein-Hawking entropy (which is obtained using semi-classical approximation \cite{1a, 2a}) become important at such short distances \cite{32, 32a, 32b, 32c, 32d}. Thus, it is important to use the Bekenstein-Hawking entropy corrected by quantum gravitational corrections to study the non-equilibrium quantum thermodynamics of black solutions at short distances \cite{j4, j1, j2}. The quantum analogue to classical work can be obtained in quantum thermodynamics using the quantum Crooks fluctuation theorem \cite{work1, work2}. Furthermore, quantum work can be related to the difference of equilibrium free energies using the Jarzynski equality, which is obtained from quantum Crooks fluctuation theorem \cite{eq12, eq14}. As it is possible to obtain equilibrium free energies for a black hole, it has been possible to investigate the quantum work done by a black hole during its evaporation \cite{j1}. Even though heat (represented in black holes by Hawking radiation) cannot be represented by an unitary information preserving process, the quantum work is represented by such a process in quantum thermodynamics \cite{12th, 12tha}. Thus, it has been argued that the black hole information paradox \cite{paradox1, paradox2} could be resolved by the use of non-equilibrium quantum thermodynamics.

The thermodynamics of black brane solutions can be studied using the original semi-classical approach \cite{bb12, bb14}.
These black brane solutions will also get modified at short distances due to quantum gravitational corrections.
The corrections to the black brane thermodynamics from such quantum gravitational corrections have been investigated \cite{j6}.
The quantum gravitational corrections to a BIon have been used to investigate the short distance modifications to their thermodynamics \cite{mir}. As F1-D3 intersection is U-dual to a system of M2-M5 branes \cite{BIon01, bion}, the quantum gravitational corrections to a BIon will be dual to gravitational corrections to a system of M2-M5 branes.
So, we will analyze the quantum gravitational corrections to a black M2-M5 bound state at short distances. We will then study the non-equilibrium quantum thermodynamics of such a brane system. It is also possible to study the phase transition in a black hole using various different information metric, such as the Ruppeiner \cite{r1, r2}, Weinhold \cite{w1, w2}, Quevedo (I and II) \cite{q1, q2}, HPEM \cite{HPEM, HPEM1, HPEM2, HPEM3}, and NTG \cite{h1, h2} metrics.
These different information metrics contain different amounts of information about the phase transition of a system.
It has been argued that the quantum gravitational corrections will modify the short distance behavior of these information metrics, and this modified behavior can be studied using an effective quantum information metric \cite{j1}. Here we will use this effective quantum information metric to study the phase transition in this system of black M2-M5 bound state.

\section{Euclidean Quantum Gravity}\label{sec1}
In this section, we will analyze the quantum gravitational corrections to the thermodynamics of a black M2-M5
bound state. The state can be obtained from the solution to the eleven-dimensional supergravity. The  metric for such a   black M2-M5 bound state is given by \cite{defo1, defo2, defo4, defo6}
\begin{eqnarray}
ds_{11}^{2}=&-&\frac{f}{(HD)^{\frac{1}{3}}}dt^{2}+\frac{H^{\frac{2}{3}}}{fD^{\frac{1}{3}}}dr^{2}+\frac{H^{\frac{2}{3}}}{D^{\frac{1}{3}}}r^{2}d\Omega_{4}^{2}\nonumber\\
&+&\frac{(dx^{1})^{2}+(dx^{2})^{2}+D\left((dx^{3})^{2}+(dx^{4})^{2}+(dx^{5})^{2}\right)}{(HD)^{\frac{1}{3}}},
\end{eqnarray}
where $H, f $ and $D$ are given by
\begin{eqnarray}
H=-1+\frac{r_{0}^{3}\sinh^{2}(\alpha)}{r^{3}},  &&
f= 1-\frac{r_{0}^{3}}{r^{3}},\nonumber\\
D= \left(\cos^{2}(\theta)+\frac{\sin^{2}(\theta)}{H}\right)^{-1}.  &&
\end{eqnarray}
This solution is  parameterized by constants
$\alpha$, $r_{0}$,  $\theta$,  which control its temperature, and brane charges.  This solution represents the M2 brane charge dissolved into the worldvolume of the M5 brane.

It is possible to obtain the equilibrium temperature $
T_{0}={3}/{(4\pi r_{0}\cosh\alpha)}$, and the equilibrium entropy $S_{0}=({\Omega_{(4)}}/ {4G})r_{0}^{4}\cosh\alpha,
$ of this solution
(with  $\Omega_{(\text{n})}=({2\pi^{{n+1}/{2}}})/{\Gamma\left({n+1}/{2}\right)}$  being the volume of the unit round $n$-sphere)   \cite{bf1}.
So, by decreasing  $r_{0}$ (size of the system), entropy decreases and  temperature increases.  As the temperature increases to a critical point, we cannot neglect the thermal fluctuations to the equilibrium thermodynamics. These thermal fluctuations are related to quantum fluctuations in the geometry of this system. Thus, we need to consider quantum gravitational corrections to the geometry of black M2-M5 bound state. This can be done using the Euclidean quantum gravity \cite{hawk}.  In Euclidean quantum gravity, the temporal coordinates in the path integral is Wick rotated in the complex plane. After such a Wick rotation, the gravitational partition function for black M2-M5 bound state can be written as (with $\beta\propto1/T$)
\begin{equation}\label{PF}
Z = \int [\mathcal{D}]e^{-\mathcal{I}_\text{E}}= \int_0^\infty \rho(E)e^{-\beta E} \text{d}E,
\end{equation}
where $\mathcal{I}_\text{E}$ is the Euclidean action for the M2-M5 bound state \cite{defo1, defo2, defo4, defo6}. Now, we can perform an inverse Laplace transform to obtain the density of states $\rho (E)$ for this system
\begin{equation} \label{CompInt}
\rho(E) =\frac{1}{2 \pi i} \int^{a+ i\infty}_{a -i\infty}e^{S(\beta)}\text{d}\beta.
\end{equation}
The entropy $S(\beta)$ for this system can be expressed in terms of this partition function as
$ S (\beta)=\beta  E+\ln Z.
$, where $E$ is the total energy of this system.
This expression for the total entropy holds at any given temperature. We define the equilibrium temperature as $T_{0}=1/\beta_0$, and the corresponding entropy as $S_{0}=S(\beta_{0})$. We can use the method of steepest decent to evaluate the integral around the saddle point $\beta_0$.  Using this approximation, it can be observed that $[\partial S(\beta)/\partial\beta]_{\beta=\beta_0}$ vanishes, and we obtain the equilibrium relation
$E=-[\partial\ln Z(\beta)/\partial\beta]_{\beta=\beta_0}$ \cite{32, 32a, 32b, 32c, 32d}. The entropy $S(\beta)$ can be expanded around the equilibrium temperature $\beta_0$ as
\begin{equation}\label{a1}
S(\beta)=S_0+\frac{1}{2}(\beta-\beta_0)^2 \left[\frac{\partial^2 S(\beta)}{\partial \beta^2 }\right]_{\beta=\beta_0}+\cdots.
\end{equation}
As the first term, $S_0=S(\beta_0)$ is the equilibrium entropy, and the second term is the first order correction to this equilibrium entropy. Now, restricting this expansion to such first order corrections, we can write
\begin{equation}
\rho(E) = \frac{e^{S_{0}}}{\sqrt{2\pi}} \left\{\left[\frac{\partial^2 S(\beta)}{\partial \beta^2 }\right]_{\beta = \beta_0}\right\}^{- \frac{1}{2}},
\end{equation}
for $[\partial^2 S(\beta)/\partial\beta^2]_{\beta=\beta_0}>0$. Thus, we observe that the corrections to the equilibrium entropy have to be proportional to $ [\partial^2 S(\beta)/\partial\beta^2]_{\beta=\beta_0} $ \cite{32, 32a, 32b, 32c, 32d}.

The thermal fluctuations can be expressed as a function of the original equilibrium temperature  $\beta_0$ for any equilibrium system, if we consider the fluctuations are close to the equilibrium \cite{fluc0, fluc1, fluc2, fluc4, 32}. However, when the microstates for the theory can be expressed in terms of a conformal field theory, the thermal fluctuations around the equilibrium can be expressed in terms of the equilibrium temperature  $\beta_0$, and the equilibrium entropy     $S_0$  \cite{32, 32a, 32b, 32c, 32d}. Such corrections have been studied for various black holes, such as black holes in five-dimensional minimal supergravity black hole \cite{grav2}, Bardeen regular black holes \cite{grav4}, Kerr-Sen-AdS black hole \cite{grav4a}, Gibbons-Maeda-Garfinkle-Horowitz-Strominger black hole \cite{grav5},
Black string \cite{grav6} and black Saturn \cite{grav7}.
Such corrections have also  been observed for a BIon \cite{mir}. As this solution representing an F1-D3 intersection is U-dual to a system of M2-M5 branes \cite{BIon01, bion}, this motivates the study of such corrections for a system of M2-M5 branes. Now the dynamics of M2-branes \cite{blg, blg1, abjm, abjm1}, M5-branes  \cite{m51}, and M2 branes ending on M5 branes \cite{wzw1, wzw2} can be described by a superconformal field theory.
It is possible to obtain the explicit expression for these thermal fluctuations as these black M2-M5 bound states will be  dual to a superconformal field theory.
The Cardy formula can be obtained from modular invariance of such a conformal field theory \cite{temp, temp0}, and it is possible to generalize Cardy formula to higher dimensions \cite{temp1, temp2, temp4, temp6}. In fact, it has been argued that by generalization the arguments based on the modular invariance of such a conformal field theory \cite{temp}, it is possible to obtain the dependence of $S(\beta)$ on the $\beta$ \cite{32, 32a, 32b, 32c, 32d}.
Thus, using from the observation that the microstates of this system can be described by a superconformal field theory, and then using the general arguments based on the modular invariance of such a superconformal field theory \cite{32, 32a, 32b, 32c, 32d}, the dependence of $S(\beta)$ on the $\beta$ can be expressed as $S(\beta) = a \beta^m + b \beta^{-n},$ with  $m, n,  a/b >0$. This function has an extremum at the equilibrium temperature, $\beta_0 = (nb/am)^{1/m+n}$ \cite{32, 32a, 32b, 32c, 32d}. Using this equilibrium temperature, it can be observed that the first order correction around the equilibrium given by $[\partial^2 S(\beta) / \partial \beta^2 ]_{\beta = \beta_0}$ have to be proportional to $\ln [S_0 \beta_0^2]$ \cite{32, 32a, 32b, 32c, 32d}. The temperature of the Hawking radiation  $T_0$ is identified with the equilibrium temperature $T_0 = 1/\beta_0$.
So, we only need the equilibrium entropy $S_0$ and equilibrium temperature $T_0$ to obtain the explicit expression for the thermal fluctuations.
The AdS/CFT correspondence has also been used to obtain such corrections to the equilibrium entropy of AdS black holes \cite{x1, x2, x3, x4}. It was observed that the entropy of such AdS black holes can be expressed as the equilibrium temperature $T_0$ and equilibrium entropy $S_0$. In fact, this is expected as based on general arguments it has been argued that for any system \cite{32, 32a, 32b, 32c, 32d}, where the microstates can be expressed in terms of a conformal field theory, the corrected entropy would be expressed in terms of the original equilibrium entropy and the original equilibrium temperature.
Thus, in AdS/CFT correspondence, the system is dual to a conformal field theory, and so it is expected corrections to the  equilibrium entropy of can be expressed in terms of such equilibrium quantities \cite{x1, x2, x3, x4}.

Such quantum corrections to the equilibrium entropy of a black hole have a universal form, and are expressed as logarithmic function of the area of horizon \cite{l1, l2, l3,   l5, l6}. However, the numerical value of the coefficient for such correction terms is model dependent
\cite{l1, l2, l3,  l5, l6}. As this coefficient is model dependent, we will keep it as a free parameter $\kappa$ in this paper \cite{ y0, y1, y3, y4, y2}.  Using this free parameter, we can write the entropy of a black M2-M5
bound state at short distances as
\begin{equation}\label{S}
S=\frac{\Omega_{\left(4\right)}}{4G}r_{0}^{4}\cosh\alpha-\kappa \ln\left(3\sqrt{\Omega_{\left(4\right)}}r_{0}\right)+\kappa \ln\left(8\pi \sqrt{G\cosh\alpha}\right).
\end{equation}
Now when $r_0$ is large, we can neglect the thermal fluctuations, at such a scale we can effectively set $\kappa \to 0$. In this limit, we obtain the equilibrium expressions \cite{bf6, defo1}. At short distances, we cannot neglect thermal fluctuations, and have to consider $\kappa \neq 0 $. The M5-brane and M2-brane charges depend on $\theta$ as $ Q_5 = \cos \theta Q, \, Q_2 = -\sin \theta Q $ and chemical potentials depend on $\theta$ as $\Phi_5 = \cos \theta \Phi, \, \Phi_2 = - \sin \theta \Phi$. However, they are fixed to be constants, and so the equilibrium entropy is independent of $\theta$ \cite{bf6, defo1}. Thus, we expect that same behavior to hold even after the thermal fluctuations are considered.
For the perturbative expansion to be valid, the leading order corrections to the equilibrium entropy have to be less than the original equilibrium entropy $S_0> \ln [S_0 T_0^2]$ \cite{y0,y1, y3, y4}. In fact, when $r_0$ is large, and the temperature is small, and $S_0>>\ln [S_0 T_0^2]$. At this scale, the thermal fluctuations can be neglected, and the system can be investigated using equilibrium thermodynamics. So, at such large scales, the thermodynamics of the M2-M5 brane system has been analyzed using its equilibrium entropy \cite{defo1, defo2, defo4, defo6}.
However, for a smaller scale, the temperature of the M2-M5 brane system increases, and this, in turn, increases thermal fluctuations. At sufficiently small scales, the thermal fluctuations become large enough such that they cannot be neglected, but still remain less than the equilibrium entropy, $S_0>\ln [S_0 T_0^2]$. At such a scale, we cannot neglect the thermal fluctuations and use equilibrium thermodynamics to analyze such a system. In fact, on such a small scale, we have to use non-equilibrium quantum thermodynamics to investigate the behavior of black holes \cite{j4, j1, j2}. It is possible to relate the non-equilibrium quantum thermodynamics to equilibrium free energy between two states of the system \cite{work1, work2}. However, this analysis founders on such a small scale, where the thermal fluctuations become of the same order as the original equilibrium entropy $S_0\sim \ln [S_0 T_0^2]$. The range of $\kappa$ is fixed such that the contributions from $\kappa$ term only become important on such a small scale. Thus, the range for the numerical value of the coefficient is taken as $\kappa = [0,1]$ \cite{ y0, y1, y3, y4, y2}. We can now investigate the behavior of the system for these different values of $\kappa$. We can use this corrected entropy to investigate the corrections to the thermodynamical properties of this system.

\section{ Quantum Work}\label{subr3}
 We have already obtained the quantum gravitational corrections to a black M2-M5 bound state at short distances. At such short distances, it is not possible to analyze the system as an equilibrium system, and have to use non-equilibrium quantum thermodynamics to study it.
 It is important to obtain quantum work to understand the behavior of the system described by non-equilibrium quantum thermodynamics \cite{adba0, adba1, adba2, adba4}.
 This can be done by using the Jarzynski inequality, which can be obtained from the quantum Crooks fluctuation theorem \cite{work1, work2}. Using the Jarzynski inequality, it is possible to relate the non-equilibrium quantum work done on a system between two states, to the difference between the equilibrium free energies of those two states. Now the quantum work done during evaporation of a black M2-M5 bound state can be obtained from the difference between the two states of such a black brane system  \cite{j1}.   This was done using non-equilibrium thermodynamics of black holes \cite{rz12, rz14}.
 The number of microstates of a black M2-M5 bound state changes, as it evaporates. Now let the black M2-M5 bound state with initial microstates  $\Omega_2$ evaporate to a  black brane with microstates  $\Omega_1$.  The non-equilibrium quantum thermodynamics can be used to obtain the quantum work done between these two states \cite{rz12, rz14}.

The microstates of a system  are related to its  entropy, and so a change in the microstates of this  black M2-M5 bound state will also change its entropy. Furthermore, as we are investigating the system at short distances, we need to use the   quantum gravitational corrected  entropy of a black M2-M5 bound state. It is possible to  express the  change in this  corrected entropy  as $r_0$ changes from an initial value of $r_{01}$ to a final value of $r_{02}$ as
\begin{eqnarray}\label{delta S}
\Delta S&=&S(r_{02}, \alpha_{2})-S(r_{01}, \alpha_{1})\nonumber \\
&=&\frac{\Omega_{\left(4\right)}}{4G}r_{02}^{4}\cosh\alpha_{2}-\kappa \ln\left(3\sqrt{\Omega_{\left(4\right)}}r_{02}\right)+\kappa \ln\left(8\pi l\sqrt{G\cosh\alpha_{2}}\right)\nonumber\\
&-&\frac{\Omega_{\left(4\right)}}{4G}r_{01}^{4}\cosh\alpha_{1}+\kappa \ln\left(3\sqrt{\Omega_{\left(4\right)}}r_{01}\right)-\kappa \ln\left(8\pi l\sqrt{G\cosh\alpha_{1}}\right).
\end{eqnarray}
The change in the corrected entropy can be used to obtain a change in the internal energy. We observe that it is important to incorporate the effects of quantum corrections  at short distances in quantum thermodynamics. In fact, we can obtain the internal energy of the system corrected by quantum gravitational corrections as
\begin{eqnarray}\label{EM}
E&=&\int T_{0}dS \nonumber\\ &=&
\frac{1}{4\pi}\frac{\Omega_{\left(4\right)}}{G}r_{0}^{3}+\frac{9\kappa\sqrt{\Omega_{\left(4\right)}}}{4\pi r_{0} \cosh\alpha}.
\end{eqnarray}
Here we have used the corrected entropy and the original equilibrium temperature $T_0$, as the thermal fluctuations are expressed as a function of the original equilibrium temperature \cite{32, 32a, 32b, 32c, 32d}.

\begin{figure}[h!]
 \begin{center}$
 \begin{array}{cccc}
\includegraphics[width=75 mm]{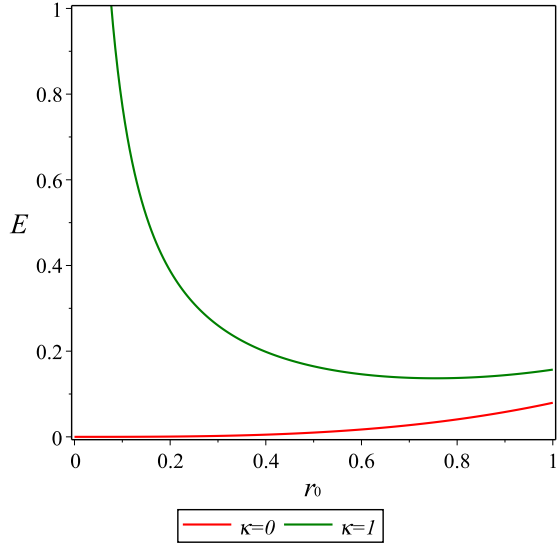}
 \end{array}$
 \end{center}
\caption{Internal energy in terms of $r_{0}$ for the unit values of the parameters.}
 \label{fig2}
\end{figure}

We can observe from Fig. \ref{fig2} that original internal energy ($\kappa =0$) increases with $r_{0}$. However, in presence of quantum fluctuations the internal energy decreases with $r_{0}$ at short distances. At the large distances, the quantum corrections do not change the behavior of the internal energy. Now we can express the change in the internal energy of the system as
\begin{eqnarray}\label{delta E}
\Delta E&=& E(r_{02}, \alpha_{2})-E(r_{01}, \alpha_{1}),
\nonumber \\
&=&\frac{1}{4\pi}\frac{\Omega_{\left(4\right)}}{G}r_{02}^{3}+\frac{9\kappa\sqrt{\Omega_{\left(4\right)}}}{4\pi r_{02} \cosh\alpha_{2}}
-\frac{1}{4\pi}\frac{\Omega_{\left(4\right)}}{G}r_{01}^{3}-\frac{9\kappa\sqrt{\Omega_{\left(4\right)}}}{4\pi r_{01} \cosh\alpha_{1}}.
\end{eqnarray}
At large values of $r_0$, the number of microstates is also very large, and the temperature of the system is very small. So, the rate of change of the microstates is much smaller than the total microstates, and hence we can neglect the change in the change in the area of the horizon. Thus, at such scales, most of the internal energy is lost through Hawking radiation, and we can neglect quantum work done by the system.

However, at short distances, quantum work cannot be neglected, and we have to use quantum thermodynamics to calculate the amount of quantum work done by this system as it evaporates from $E_1$ to $E_2$.
At such scales, the change in the total internal energy of the system can be expressed in terms of the energy spend in doing quantum work between two states, and the energy radiated in the Hawking radiation. Thus, if $\langle W \rangle$ is the average quantum work done by this system (between two states), and $Q$ is the total heat energy radiated by the Hawking radiation, then we can write \cite{10th}
\begin{equation}
\Delta E = Q - \langle W \rangle
\end{equation}
It is possible to use the Jarzynski inequality to estimate the quantum work done as this black M2-M5 bound state as they evaporate from $\Omega_1$ to $\Omega_2$    \cite{work1, work2}. This can be done by relating the quantum work to the  difference of the equilibrium free energies between two states of the system \cite{eq12, eq14}
\begin{equation}
\langle e^{-\beta W} \rangle = e^{\beta \Delta F}.
\end{equation}

The system can be seen to have a dual description in terms of a superconformal field theory, and a black geometry.  The black geometry of M2-M5 branes emit Hawking radiation. The  energy in the  Hawking radiation is represented by heat, and heat is not an unitary information preserving process in quantum thermodynamics \cite{12th, 12tha}.  However, as this system has a dual description in terms of a superconformal field theory, it becomes important to analyze its short distance behavior. This is because from the superconformal side of the duality, this system is represented by an unitary superconformal field theory. We thus apply the formalism of quantum non-equilibrium thermodynamics to it black geometry, and calculate quantum work in it.
The  quantum work is  represented  by an unitary information preserving process in quantum thermodynamics \cite{12th, 12tha}. Thus, it seems that the way to reconcile the dual description of M2-M5 branes  would be to use such non-equilibrium quantum thermodynamics. We would like to point out that such analysis has been recently been applied to rotating  black hole   \cite{j1}. In this analysis, it was demonstrated how quantum work can be obtain from this black hole, and how the formalism of non-equilibrium quantum  thermodynamics enables such a black holes  to lose its mass through an unitary process. As the equilibrium  thermodynamics of
a system of M2-M5 branes has been thoroughly studied  \cite{bf6, defo1}, it is interesting to apply the formalism of quantum thermodynamics and analyze quantum work for it. It has been argued that information can leak out of a black hole during last stages of its evaporation from such an unitary information preserving process \cite{j1}. Thus, it might be possible to resolve the   information loss paradox  in such a system using non-equilibrium quantum thermodynamics \cite{paradox1, paradox2}, which is expected due to its dual description in terms of superconformal field theory. However, on such a small scale, we have to analyze the system using   non-equilibrium  quantum thermodynamics. As  quantum work, which is represented by an unitary  information preserving process, becomes important at short distances, it is possible that black hole information paradox can be resolved by the use of  non-equilibrium quantum thermodynamics. This can be done by using the relation between  non-equilibrium quantum thermodynamics  and information theory \cite{info1, info2}. It could be possible that the absence of information paradox   in gauge gravity duality occurs due to   this  unitary  information preserving process  \cite{paradox4, paradox5}.
Now, we can obtain the quantum work using  quantum corrected free energy.  We can write the difference between  free energy,  which has been corrected by  quantum gravitational corrections as
\begin{eqnarray}\label{Delta F}
\Delta F&=&\frac{\Omega_{\left(4\right)}}{4\pi G}(r_{02}^{2}-r_{01}^{2})-\frac{3\Omega_{\left(4\right)}}{16\pi G}(r_{02}^{3}-r_{01}^{3})
+\frac{45\kappa\sqrt{\Omega_{\left(4\right)}}}{16\pi }\left(\frac{1}{r_{02} \cosh\alpha_{2}}-\frac{1}{r_{01} \cosh\alpha_{1}}\right)\nonumber\\
&-&\frac{9\kappa G}{4\pi\sqrt{\Omega_{\left(4\right)}}}\left(\frac{\ln{(3\sqrt{\Omega_{\left(4\right)}}r_{02})}}{r_{02}^{5}\cosh^{2}\alpha_{2}}
-\frac{\ln{(3\sqrt{\Omega_{\left(4\right)}}r_{01})}}{r_{01}^{5}\cosh^{2}\alpha_{1}}\right)
+\frac{3\kappa }{4\pi}\left(\frac{\ln{(3\sqrt{\Omega_{\left(4\right)}}r_{02})}}{r_{02}\cosh\alpha_{2}}
-\frac{\ln{(3\sqrt{\Omega_{\left(4\right)}}r_{01})}}{r_{01}\cosh\alpha_{1}}\right)\nonumber\\
&-&\frac{3\kappa }{4\pi G}\left(\frac{\ln{(8\pi l\sqrt{G \cosh\alpha_{2}})}}{r_{02}\cosh\alpha_{2}}
-\frac{\ln{(8\pi l\sqrt{G \cosh\alpha_{1}})}}{r_{01}\cosh\alpha_{1}}\right)\nonumber\\
&+&\frac{3\kappa^{2}G }{4\pi \Omega_{\left(4\right)}^{\frac{3}{2}}}\left(\frac{\ln{(8\pi l\sqrt{G \cosh\alpha_{2}})}}{r_{02}^{5}\cosh^{2}\alpha_{2}}
-\frac{\ln{(8\pi l\sqrt{G \cosh\alpha_{1}})}}{r_{01}^{5}\cosh^{2}\alpha_{1}}\right).
\end{eqnarray}

The Jensen inequality can be used to express average of exponential of quantum work in terms of the  exponential  of the average of quantum work for this system as
$ e^{\langle -\beta {W} \rangle } \leq \langle e^{-\beta W} \rangle
$. Thus, an inequality for the  quantum  work done can be obtained using the Jensen inequality. The   Jarzynski
inequality relates the average quantum work to the difference between equilibrium free energies $\langle  W \rangle \geq \Delta F$ \cite{j1}. So,  neglecting the second order correction, we can write  the minimum value for average quantum work $\langle  W \rangle_{min} = \Delta F $ as
\begin{equation}\label{W}
\langle  W \rangle_{min} =  \langle  W \rangle_{0}+\kappa \langle  W \rangle_{1},
\end{equation}
where $\langle  W \rangle_{0}$ is the original minimum quantum work  and $\langle  W \rangle_{1}$ is the quantum gravitational correction to the minimum quantum work. We can explicitly write the expressions for the  minimum quantum work and quantum gravitational correction to the minimum quantum work as
\begin{eqnarray}\label{W0}
\langle  W \rangle_{0}&=&\frac{\Omega_{\left(4\right)}}{4\pi G}\left[\frac{3}{4}(r_{02}^{3}-r_{01}^{3})-r_{02}^{2}+r_{01}^{2}\right],
\\
\langle  W \rangle_{1}&=&\frac{9 G}{4\pi\sqrt{\Omega_{\left(4\right)}}}\left(\frac{\ln{(3\sqrt{\Omega_{\left(4\right)}}r_{02})}}{r_{02}^{5}\cosh^{2}\alpha_{2}}
-\frac{\ln{(3\sqrt{\Omega_{\left(4\right)}}r_{01})}}{r_{01}^{5}\cosh^{2}\alpha_{1}}\right)\nonumber\\
&-&\frac{3}{4\pi}\left(\frac{\ln{(3\sqrt{\Omega_{\left(4\right)}}r_{02})}}{r_{02}\cosh\alpha_{2}}
-\frac{\ln{(3\sqrt{\Omega_{\left(4\right)}}r_{01})}}{r_{01}\cosh\alpha_{1}}\right)\nonumber\\
&+&\frac{3}{4\pi G}\left(\frac{\ln{(8\pi l\sqrt{G \cosh\alpha_{2}})}}{r_{02}\cosh\alpha_{2}}
-\frac{\ln{(8\pi l\sqrt{G \cosh\alpha_{1}})}}{r_{01}\cosh\alpha_{1}}\right)\nonumber\\
&-&\frac{45\sqrt{\Omega_{\left(4\right)}}}{16\pi }\left(\frac{1}{r_{02} \cosh\alpha_{2}}-\frac{1}{r_{01} \cosh\alpha_{1}}\right).
\end{eqnarray}
It may be observed that the minimum value for the quantum work changes due to quantum gravitational effects at short distances. However, this can again be neglected at large distances.



The partition function for this system can be obtained from its microstates. As a conformal field theory is dual to this system, we can estimate its microstates, and obtain its partition function from the dual theory. Now we assume that these microstates change from $\Omega_1$ to $\Omega_2$ during the process of evaporation. This change in the microstates will also change the partition function changes for the system. Let us assume that the initial partition function is   $Z_1 [\Omega_1]$, and it changes to a final partition function  $Z_2[\Omega_2]$. We can relate these partition functions for this system using the Jarzynski inequality as
\cite{eq12, eq14}
\begin{equation}
\langle e^{-\beta  W  } \rangle \nonumber=\frac{Z_2}{Z_1}
\end{equation}

So, the relative weights of the partition functions  $Z_2/Z_1$ can be expressed in terms of average quantum work done between the states $\Omega_1$ and $\Omega_2$.  As the average quantum work can be related to the difference between the equilibrium free energies $\Delta F$ of these two states, we can also write the relative weights of the partition function as $  \exp({\beta \Delta F}) ={Z_2}/{Z_1}$.

\section{Quantum Effective Informational Metrics}\label{subr2}

\begin{figure}[h!]
 \begin{center}$
 \begin{array}{cccc}
\includegraphics[width=75 mm]{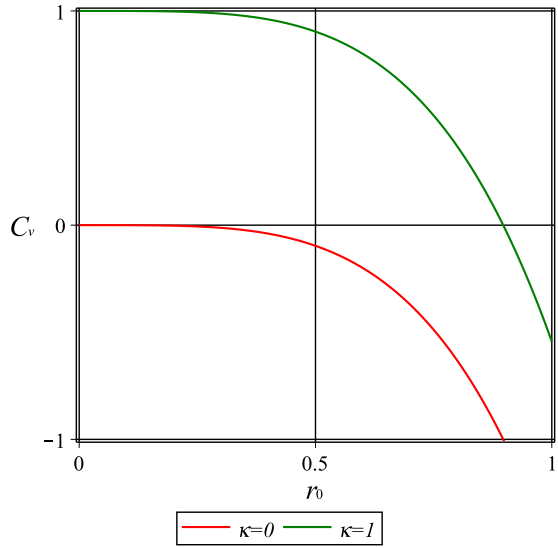}
 \end{array}$
 \end{center}
\caption{Specific heat at constant volume in terms of $r_{0}$ for the unit values of the parameters.}
 \label{fig1}
\end{figure}

It is expected that the quantum corrections to the black M2-M5 bound state will modify the thermodynamic stability of this solution. Such modification to the thermodynamic behavior from quantum corrections has been studied for various black objects \cite{40a, 40b, 40c, 40d}.
The stability of this system can be investigated using its specific heat $
C =T_{0}({\partial S}/{\partial T_{0}})
$. Thus, we can express the quantum gravitationally corrected specific heat of a black M2-M5
bound state as
\begin{equation}\label{C0}
C =-\frac{\Omega_{\left(4\right)}}{G}r_{0}^{4}\cosh\alpha + \kappa
\end{equation}
Here we observe that when $r_0$ is very large, such that $\kappa << {\Omega_{\left(4\right)}}r_{0}^{4}\cosh\alpha/G$, we can analyze the stability of the system using its equilibrium description. However, at smaller values of $r_0$, when the contribution from thermal fluctuations cannot be neglected, and still remain small as compared to equilibrium contributions, $\kappa < {\Omega_{\left(4\right)}}r_{0}^{4}\cosh\alpha/G$, we can use this expression for the corrected specific heat. Nonetheless, on very small scales, when the contributions become of the same order as the equilibrium contributions $\kappa\sim {\Omega_{\left(4\right)}}r_{0}^{4}\cosh\alpha/G$, this approximation breaks down, and we cannot analyze the system using small fluctuations around an equilibrium. It is clear that the specific heat is negative in absence of quantum fluctuations. Hence, a black M2-M5
bound state is thermodynamically unstable. Having said that, in presence of quantum fluctuations ($\kappa\neq0$), this system becomes stable at smaller $r_{0}$. We observe that the quantum gravitational corrections modify the thermodynamic stability of this black brane configuration at short distances, and it becomes stable at short distances due to quantum fluctuations. We also observe there that the thermodynamic behavior of the system does not change at large distances. This is expected, as the change occurs from quantum fluctuations, which can be neglected at such large distance scales.
This is illustrated by plots of Fig. \ref{fig1}.\\

The thermodynamic stability of black holes can be investigated using information geometry \cite{1111, 2222}. The divergences of the Ricci scalar of the information geometry has been used to study the phase transitions in such thermodynamic systems \cite{point1, point2, point4, point5, badpa, 3dmpl}. The form of the Ricci scalar for a black hole depends on the specific form of the information metric used in the information geometry. Different amounts of information can be extracted from different information-theoretical metrics. The Ruppiner metric has been used to investigate the phase transition in a charged Gauss-Bonnet AdS black holes \cite{12ra}. It has also been observed that the Weinhold metric can describe the phase transition in a Park black hole \cite{12rw}.
Even though the phase transition in an AdS black hole with a global monopole can be described by neither Weinhold nor Ruppiner metrics \cite{12qb}, it is possible to study its phase transition using Quevedo and HPEM metrics \cite{12qb}. Furthermore, both the Ruppiner and HPEM metrics can be used to study the phase transition of a black hole surrounded by the perfect fluid in Rastall theory \cite{rh12}. Here HPEM metrics contains more information about the phase transition than the Ruppiner metric. The Weinhold and Quevedo metrics cannot be used to study the phase transition of this system. As different information-theoretical metrics can provide different amounts of information about the phase transition of a black object, it is important to use different information-theoretical metrics to properly analyze such a phase transition.

\begin{figure}[h!]
 \begin{center}$
 \begin{array}{cccc}
\includegraphics[width=60 mm]{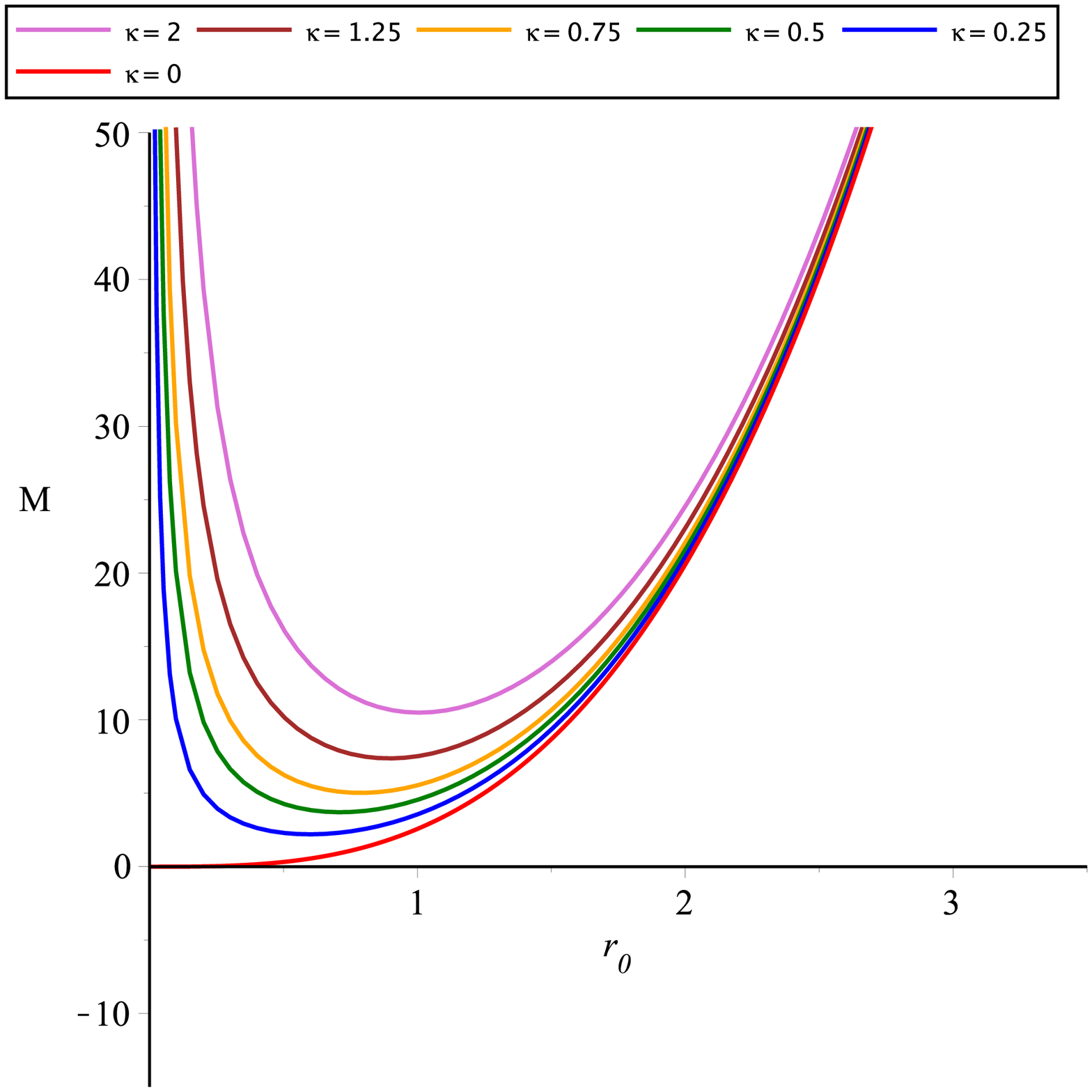}\includegraphics[width=60 mm]{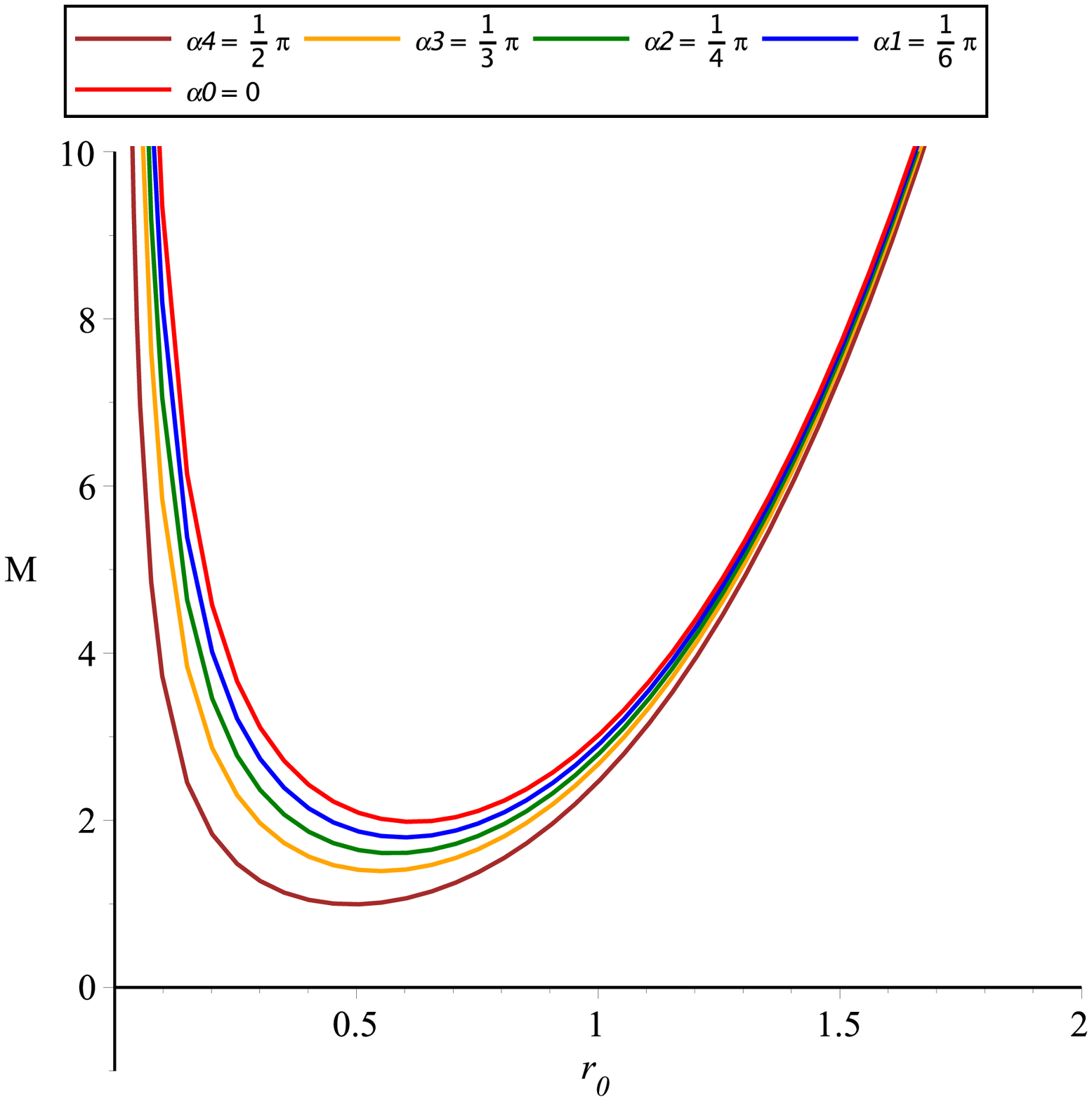}
 \end{array}$
 \end{center}
\caption{Variations of quantum mass with  horizon radius $ r_{0}$ (a) left: $\alpha=\frac{\pi}{6}$; (b) right: $\kappa=0.25$.}
 \label{pic:M}
\end{figure}

At short distances, the effects of quantum gravitational corrections modify the thermodynamic behavior of the system. So, they will also modify the information-theoretical geometry associated with its thermodynamics. Now it is known that the
 information-theoretical metrics are constructed using the mass of the black object. Thus, the quantum gravitational corrections to the information-theoretical geometry can be constructed by using novel quantum mass for the system. We can use the equilibrium thermodynamics for a system of M2-M5 brane \cite{bf6, defo1} to define a mass for it, when $r_0$ is large \cite{dolan}, which can then be used to analyze different information-theoretical metrics. However, we have to use the expression for thermal fluctuations to define a novel quantum mass for the system. Here we expect that in the limit, $\kappa \to 0$, this quantum mass reduces to the original mass for the system. This can be done by calculating internal energy $E$ for the black hole solution, and using it to define the mass at large $r_0$ as $M\equiv E$ \cite{dolan}. Such a mass has been used to different analyzed information geometries and the phase transitions in various black holes \cite{point1, point2, point4, point5, badpa, 3dmpl}. In fact, it has been used to construct Ruppeiner \cite{r1, r2}, Weinhold \cite{w1, w2}, Quevedo (I and II) \cite{q1, q2}, HPEM \cite{HPEM, HPEM1, HPEM2, HPEM3}, and NTG \cite{h1, h2} metrics. Now it was observed that internal energy given in Eq. (\ref{EM}) is corrected due to thermal fluctuations, which were calculated using Euclidean quantum gravitational partition function. So, using the expression for this corrected internal energy from Eq. (\ref{EM}), we can define a novel quantum mass for the system. Such a novel quantum mass can then be used to analyze the effects of quantum corrections to the information geometries associated with this solution \cite{j1}. Thus, using $M\equiv E$, we can define the novel quantum mass for this solution as
\begin{eqnarray}
M&=&
\frac{1}{4\pi}\frac{\Omega_{\left(4\right)}}{G}r_{0}^{3}+\frac{9\kappa\sqrt{\Omega_{\left(4\right)}}}{4\pi r_{0} \cosh\alpha} \nonumber \\
&=&  \frac{1}{8}\dfrac{\sqrt{2}\Omega_{\left(4\right)}\bigg(9 \kappa\sqrt{\Omega_{\left(4\right)}}+4 S_{0}\bigg)}{\pi\bigg(G \Omega_{\left(4\right)}^{3}\cosh^{3}(\alpha) S_{0}\bigg)^{\frac{1}{4}}}.
\end{eqnarray}
The behavior of this novel quantum mass with $ r_{0} $ has been plotted in Fig.~\ref{pic:M}. It can be observed from Fig.~\ref{pic:M} that the quantum mass has a lower bound. In the left plot of Fig.~\ref{pic:M}, the variation of quantum mass with $\kappa $ is plotted. In the right plot of Fig.~\ref{pic:M}, the variation of quantum mass with the parameter $\alpha$ is plotted. The value of this lower bound increases by increasing the value of $\kappa$ (see Fig.~\ref{pic:M} (a)), and decreases by increasing the value of $\alpha$ (see Fig.~\ref{pic:M} (b)). We have denoted $\kappa = 0$ by solid red line on Fig.~\ref{pic:M} (a). Here we observe that it vanishes at $r_{0}=0$. However, for any other value of $\kappa$, there is a finite minimum bound for this quantum mass.

\begin{figure}[h]
	\centering
	\subfigure[]{
		\includegraphics[width=0.4\textwidth]{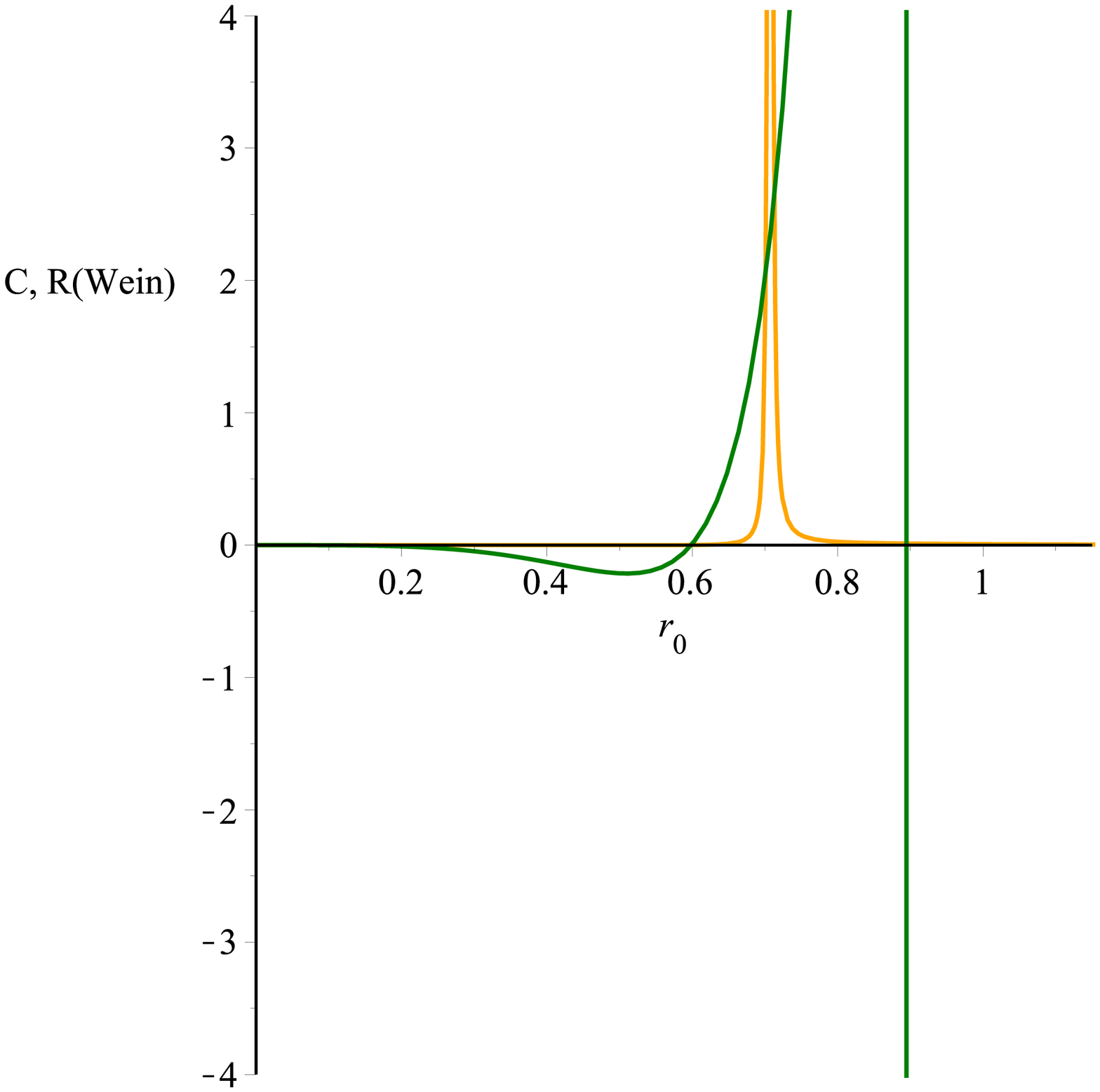}
	}
	\subfigure[]{
		\includegraphics[width=0.4\textwidth]{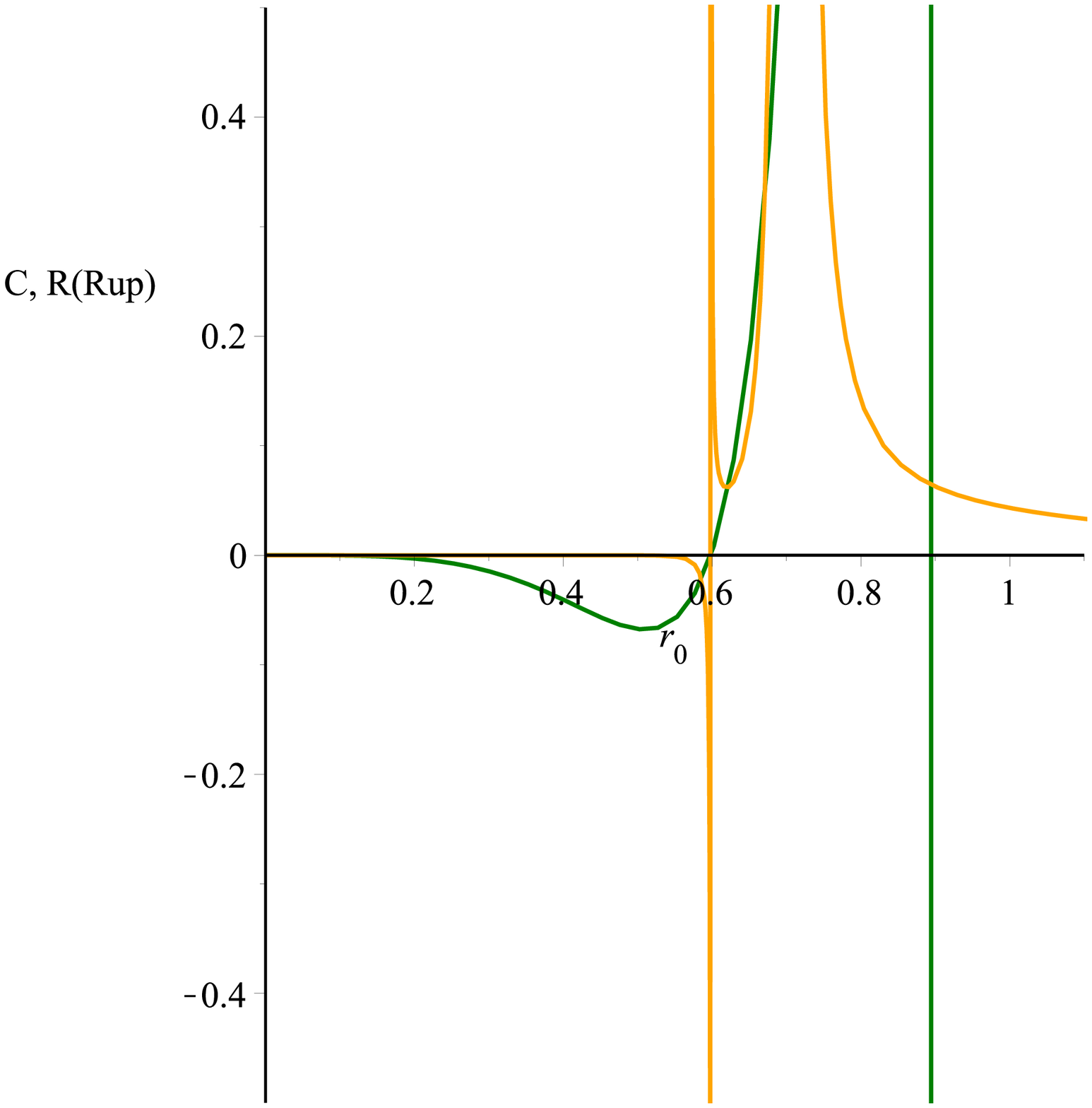}
    }
	\subfigure[]{
		\includegraphics[width=0.4\textwidth]{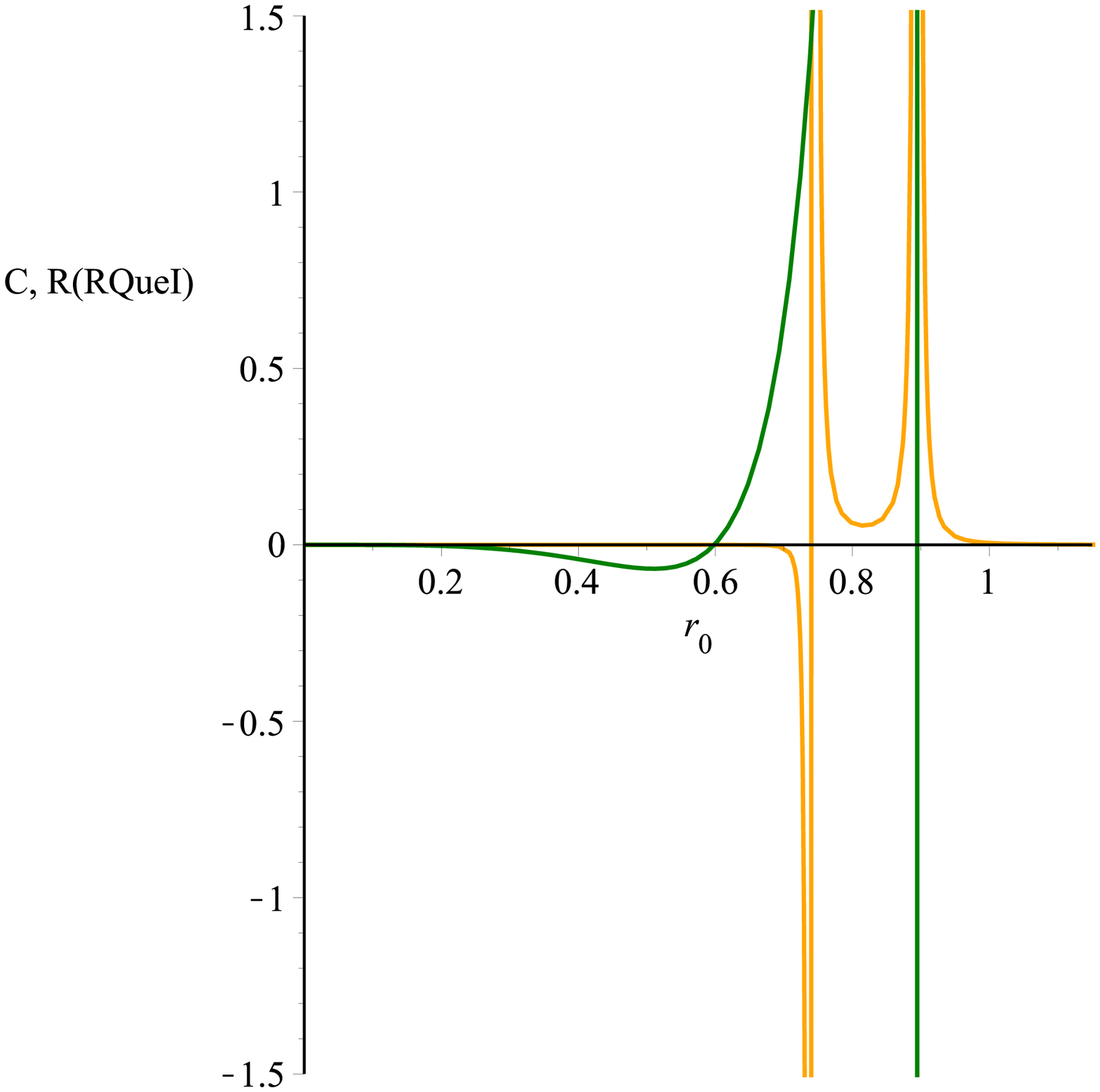}
	}
    \subfigure[]{
	\includegraphics[width=0.4\textwidth]{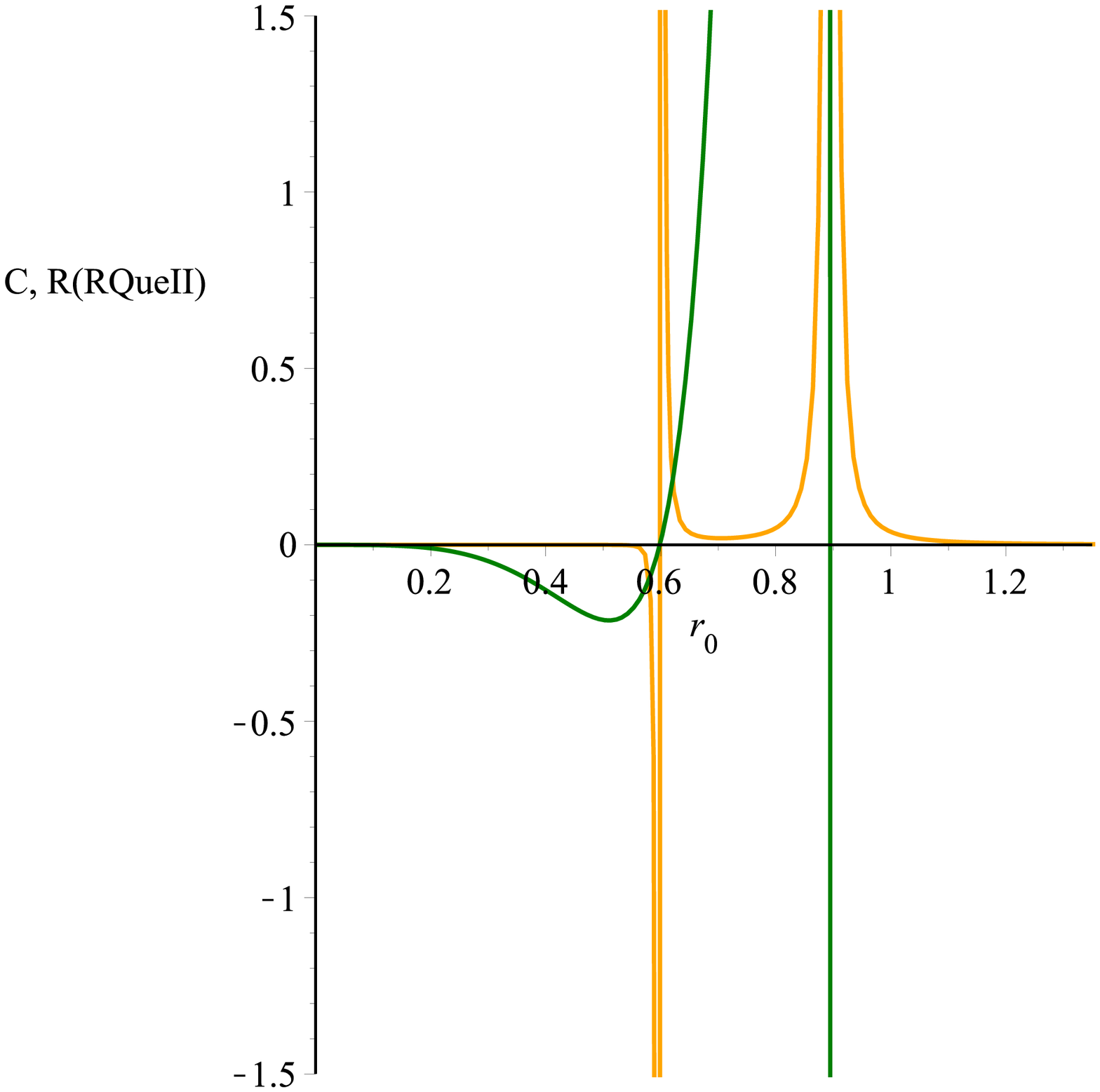}
    }
	\subfigure[]{
		\includegraphics[width=0.4\textwidth]{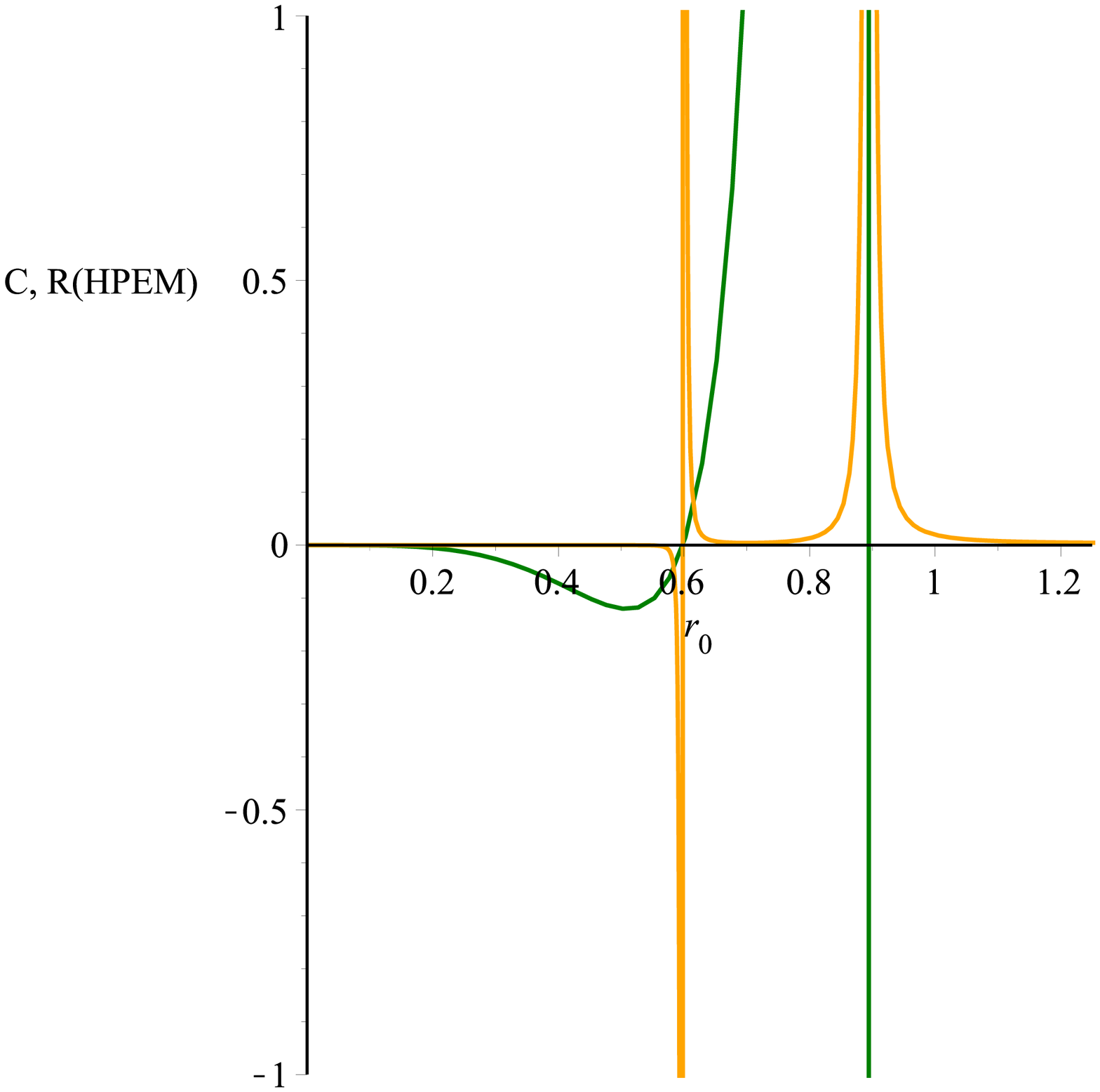}
	}
	\subfigure[]{
		\includegraphics[width=0.4\textwidth]{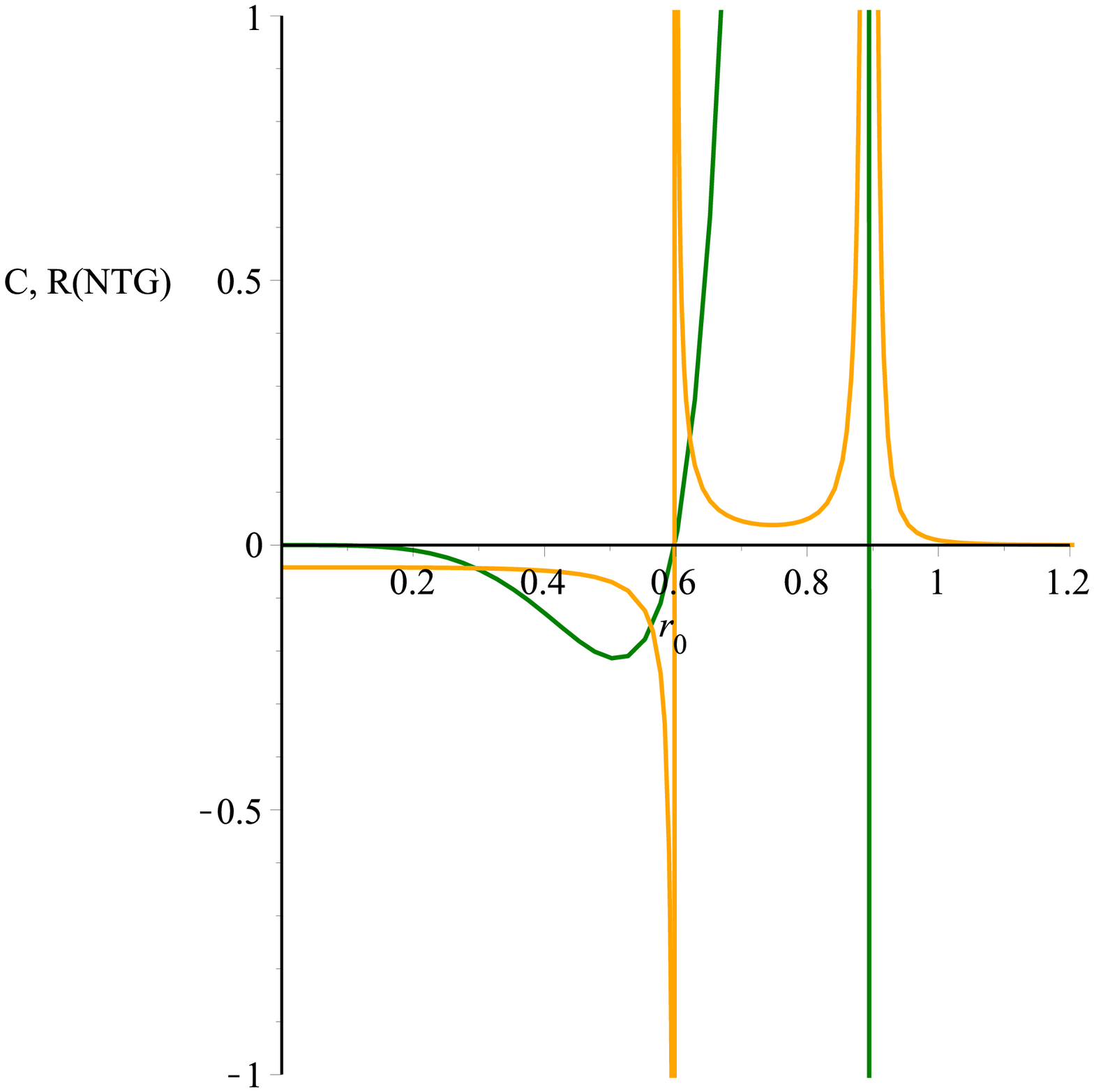}
	}
	\caption{Curvature scalar variation of Weinhold, Ruppeiner, HPEM and NTG metrics (orange line) and also heat capacity variation (Green line), in terms of horizon radius $ r_{0} $ for $ \kappa=0.25 $ and $\alpha=\frac{\pi}{6} $, for a black M2-M5
bound state.}
	\label{pic:CRWeinQueHPEMNTG}
\end{figure}

We can construct quantum corrected effective using the novel quantum mass for the system. We will use the Ruppeiner metric to analyze the phase transition in this system.
This metric is constructed using the space of thermodynamical fluctuations \cite{r1, r2}. We can obtain an effective Ruppeiner metric by incorporating the quantum gravitational corrections in the original Ruppeiner of this system. This quantum corrected effective Ruppeiner metric can be written as
\begin{equation}
ds^2=   -\dfrac{1}{T}Mg_{ab}^{R}dX^{a}dX^{b} .
\end{equation}
We can also use the Weinhold metric to analyze the phase transition of this system. The relation between Gibbs-Duhem relation and scaling of thermodynamic potentials is used to obtain this metric \cite{w1, w2}. Similar to writing the quantum corrected effective Ruppeiner metric, we can obtain a quantum corrected effective Weinhold metric as
\begin{equation}
ds^2=  Mg_{ab}^{W}dX^{a}dX^{b}.
\end{equation}
The Legendre invariant set of metrics in the phase space can be used to construct Quevedo (I and II) metrics \cite{q1, q2}. The information metrics on the space of equilibrium states can be constructed using their pullback. Now by correcting the original Quevedo (I and II) metrics by quantum gravitational corrections, we obtain quantum corrected effective Quevedo (I and II) metrics. The quantum corrected effective Quevedo I metric can be expressed as
\begin{equation}
ds^2=(SM_{S}+\alpha M_{\alpha})(-M_{SS}dS^{2}+M_{\alpha\alpha}d\alpha^{2})
\end{equation}
We can also express the quantum corrected effective Quevedo II metric as
\begin{equation}
ds^2= \quad SM_{S}(-M_{SS}dS^{2}+M_{\alpha \alpha}d\alpha^{2}).
\end{equation}
It is possible to use a different conformal factor than the Quevedo metrics to obtain the HPEM metric \cite{HPEM, HPEM1, HPEM2, HPEM3}. Now by incorporating the quantum gravitational corrections into the HPEM metric we can obtain the quantum corrected effective HPEM metrics as
\begin{equation}
ds^2=\quad \quad \dfrac{SM_{S}}{\left(\frac{\partial^{2} M}{\partial \alpha^{2}}\right)^{3}}\left(-M_{SS}dS^{2}+M_{\alpha\alpha}d\alpha^{2}\right).
\end{equation}
The NTG metric is obtained by using the Jacobean transformation to change the coordinates of the thermodynamic space \cite{h1, h2}. We can again correct it by quantum gravitational correction to obtain a quantum corrected effective NTG metric
\begin{equation}
ds^2 = \quad \quad \dfrac{1}{T}\left(-M_{SS}dS^{2}+M_{\alpha\alpha}d\alpha^{2}\right).
\end{equation}
Here for these quantum corrected effective metrics, $M_{XY}$ denotes the second-order differentiation of $M$ with respect to the thermodynamic variables $X$ and $Y$.

The variation of the scalar curvature and heat capacity for these metrics with $ r_{0} $ has been plotted in Fig.~\ref{pic:CRWeinQueHPEMNTG}. It is clear from Fig.~\ref{pic:CRWeinQueHPEMNTG} (a) that the quantum effective Weinhold metric is not able to explain the phase transitions of this system. In Fig.~\ref{pic:CRWeinQueHPEMNTG} (b) it has been demonstrated that the scalar curvature of quantum effective Ruppeiner metric only becomes singular for zero point of the heat capacity. It does not have any singular point corresponding to divergent points of the heat capacity (critical points associated with a phase transition). In Fig.~\ref{pic:CRWeinQueHPEMNTG} (c), divergence of the curvature scalar of quantum effective Quevedo case (I) formalism coincides only with the divergence point of the heat capacity, and it does not have any point corresponding to the zero point of the heat capacity. However, from Fig.~\ref{pic:CRWeinQueHPEMNTG} (d), (e), (f), we observe that the singular points of the scalar curvature for the quantum effective Quevedo case (II), quantum effective HPEM and quantum effective NTG metrics, coincide with both the zero point and the divergence points of heat capacity. Thus, in the quantum effective Quevedo case (II), quantum effective HPEM and quantum effective NTG metrics, we have a one-to-one correspondence between singularities of the scalar curvature of these metrics and phase transitions of the heat capacity. So, to investigate the phase transition of this system, we need to use the quantum effective Quevedo case (II), quantum effective HPEM and quantum effective NTG metrics. This is because they provide more information about this system than the quantum effective Weinhold, quantum effective Ruppeiner and quantum effective Quevedo case (I) metrics.
\clearpage
\section{Conclusion}\label{subrc}
In this paper, we have analyzed the quantum thermodynamics for an M2-M5 bound state. We have analyzed the quantum corrections to the thermodynamics of this system. This has been done using the Euclidean quantum gravity. Thus, we have observed that these quantum gravitational corrections produce logarithmic corrections to the entropy of this system. We have demonstrated that these logarithmic corrections, which come from the thermal fluctuations, make the system stable at the quantum scales. We obtained corrected quantum work by using the Helmholtz free energy of the system.
This was important as the system have a dual description in terms of a superconformal field theory and a black geometry. Now the black geometry of this system emits Hawking radiation, which represents heat. The heat is not an unitary information preserving process \cite{12th, 12tha}, but the system has to be represented by an unitary process due to its dual description in terms of a superconformal field theory. We thus applied the formalism of quantum non-equilibrium thermodynamics to its black geometry, and calculated quantum work in it. The advantage of using quantum work is that it is represented by unitary information preserving process in quantum thermodynamics \cite{12th, 12tha}. Thus, we proposed that to reconcile the dual description of M2-M5 branes we have to analyze it using non-equilibrium quantum thermodynamics. Furthermore, the equilibrium thermodynamics of a system of M2-M5 branes has been thoroughly studied \cite{bf6, defo1}, so it was interesting to apply the formalism of quantum thermodynamics and analyze quantum work for it. We have also explicitly analyzed the corrections to thermodynamic information geometry from quantum gravitational corrections to this system. This was done by first using the quantum gravitational corrections to the internal energy to define a novel quantum mass, and then using that novel quantum mass to correct the information geometries for this system. We also compared the relative effects of the thermal fluctuations on various different informational theoretical metrics.
It is interesting to study higher-order corrections on the entropy including an exponential term on the system considered in this paper.

Here we analyzed the quantum corrected information geometry by incorporating quantum gravitational corrections into effective quantum information-theoretical metrics. It would be interesting to analyze this information geometry corrected by quantum gravitational corrections using the AdS/CFT correspondence. As different information metrics contain different amounts of information about the system, it would be interesting to analyze their dual structure. It would also be interesting to study holographic complexity and holographic entanglement entropy for this system, with quantum gravitational corrections. It is expected that the holographic entanglement entropy would also get corrected from such quantum gravity corrections. These corrections could also modify the holographic complexity of this system.
The geometry of spacetime is an emergent structure in the Jacobson formalism. Here the geometry emerges from the thermodynamics of the system. Here we were able to analyze the quantum gravitational corrections to the thermodynamics of a black M2-M5 bound state. Thus, it is possible to use this corrected thermodynamics to obtain corrections to the metric of this system using the Jacobson formalism. We could, therefore, obtain an effective quantum corrected metric for this M2-M5 brane system. At large distances, the effect of quantum gravitational corrections could be neglected, and this effective quantum metric for the M2-M5 brane system would reduce to the usual metric for a black M2-M5 bound state. However, at short distances, the quantum gravitational corrections would modify the original metric for the M2-M5 brane system. This modification of the M2-M5 brane system by quantum gravitational effects would also modify the Raychaudhuri equation for such a system. This quantum corrected Raychaudhuri equation can be used to investigate modifications to the singularity theorems from quantum gravitational corrections. These quantum corrections to the system would also modify the geometry flow in such a system. It would be interesting to investigate such modifications to the system of the M2-M5 brane system at short distances.


\end{document}